\documentclass[aps,amsmath,amssymb,prb,showpacs,floatfix,preprint,superscriptaddress,nofootinbib]{revtex4-1}
\usepackage{bm}
\usepackage[colorlinks=true,citecolor=blue,linkcolor=black,urlcolor=blue]{hyperref}

\begin{document}

\title{Holographic duality and the resistivity of strange metals}

\author{Richard A. Davison}
\email{davison@lorentz.leidenuniv.nl}
\affiliation{Instituut-Lorentz for Theoretical Physics, Universiteit Leiden, P.~O.~Box 9506, 2300 RA Leiden, The Netherlands}
\author{Koenraad Schalm}
\email{kschalm@lorentz.leidenuniv.nl}
\affiliation{Instituut-Lorentz for Theoretical Physics, Universiteit Leiden, P.~O.~Box 9506, 2300 RA Leiden, The Netherlands}
\affiliation{Department of Physics, Harvard University, Cambridge MA 02138, USA}
\author{Jan Zaanen}
\email{jan@lorentz.leidenuniv.nl}
\affiliation{Instituut-Lorentz for Theoretical Physics, Universiteit Leiden, P.~O.~Box 9506, 2300 RA Leiden, The Netherlands}

\begin{abstract}

We present a strange metal, described by a holographic duality, which reproduces the famous linear resistivity of the normal state of the copper oxides, in addition to the linear specific heat. This holographic metal reveals a simple and general mechanism for producing such a resistivity, which requires only quenched disorder and a strongly interacting, locally quantum critical state.  The key is the minimal viscosity of the latter: unlike in a Fermi-liquid, the viscosity is very small and therefore is important for the electrical transport. This mechanism produces a resistivity proportional to the electronic entropy.

\end{abstract}

 \maketitle

\section{Introduction}

The ``strange metals'', as realised in high $T_c$ superconductors and related systems, are among the great mysteries of modern physics. \cite{Zaanen:2010yk} It is believed that these are strongly interacting ``quantum soups'' controlled by an emergent temporal scale invariance of their underlying quantum dynamics. \cite{PhysRevLett.63.1996} The general principles governing such strongly interacting quantum critical states could be captured by the holographic dualities discovered in string theory. \cite{Maldacena:1997re} However, an explanation of the iconic linear resistivity \cite{Zaanen:2010yk,PhysRevLett.63.1996} observed experimentally remains elusive. 

Holographic duality asserts that some strongly interacting quantum field theories are mathematically equivalent to certain classical theories of gravity in higher-dimensional curved spacetimes. \cite{Maldacena:1997re} This is a very useful theoretical tool as it allows us to easily calculate, in a well-controlled way, properties of certain strongly interacting, quantum critical states of matter by analysing their classical gravitational duals. Analyses of this kind have found examples of systems, controlled by quantum critical IR fixed points, with linear resistivities,\cite{Faulkner:2010zz,Donos:2012ra,Hoyos:2013eza} amongst other interesting phenomena. While interesting, many of these results appear to be highly dependent upon the microscopic details of the IR fixed point being considered and, a priori, it is not clear whether these details are generic and could be realised in real electron systems, or whether they are artefacts of the kinds of highly symmetric quantum field theories which are dual to classical theories of gravity. 

In this paper, we address these issues by describing a simple universal mechanism by which strongly interacting, quantum critical states of matter can acquire a linear resistivity, and propose that this could be at work in the strange metallic phase of the cuprates. This mechanism is at work in certain strongly interacting field theories which have holographic duals (specifically, those controlled by a locally quantum critical IR fixed point). However, it is based on general physical principles which do not rely on the existence of a dual, classical gravitational description of the state. Specifically, a theory whose IR physics is well-described by the laws of hydrodynamics with a ``minimal'' viscosity will have, when weakly coupled to random disorder, a viscous contribution to its resistivity which is proportional to its thermodynamic entropy. A state with a Sommerfeld heat capacity (entropy $S\sim T$), such as the cuprate strange metals, will therefore have a resistivity linear in temperature $T$.

The physical principles which govern the transport of charge in highly collective, charged quantum critical states of matter are rather different from those in an ordinary Fermi liquid composed of long-lived electronic quasiparticles (see references \onlinecite{Hartnoll:2007ih,Hartnoll:2008hs,Hartnoll:2012rj,Anantua:2012nj,Mahajan:2013cja,Hartnoll:2014gba} for some recent studies of systems of this type). In particular, the absence of long-lived quasiparticles means that the natural quantities to consider are the collective properties of the system: its total electrical current and momentum. In a system of this type, one intuitively expects that the intrinsic decay rate of the electrical current will generically be large. However, in the absence of a background lattice or impurities, the DC resistivity will vanish. This is because the electrical current carries momentum, which is conserved due to the translational invariance of the system, and thus it cannot decay. If we couple such a system weakly to impurities or a background lattice, such that the intrinsic interaction rate of the electronic fluid is much greater than the rate at which it interacts with the impurities or lattice, the momentum will dissipate slowly and will control the decay of electrical current at late times. The DC resistivity is therefore controlled by the decay rate of the total momentum of the state, in contrast to normal metals in which the decay rate of quasiparticle excitations plays a similar role.

We show that, if the collective state obeys the laws of hydrodynamics, the momentum dissipation rate in the presence of impurities is determined by the transport coefficients of the liquid, including its viscosity, which depend upon the microscopic details of the system. It is natural to expect the viscosity of a fluid arising from a strongly interacting quantum critical system to be small because, in a kinetic theory of quasiparticles, the viscosity is proportional to the mean free time between quasiparticle collisions. This can be made much more precise: the ``almost perfect'' fluids formed by strongly interacting, quantum critical systems are expected to have a viscosity of the order of the entropy density of the state, in units of $\hbar/k_B$.\cite{Sachdev,Kovtun:2004de,Iqbal:2008by} This is a universal feature of systems with holographic duals, and has been measured experimentally in strongly interacting, collective systems without such duals.\cite{Schafer:2009dj,Heinz:2013th} In this scenario, there is a universal viscous contribution to the momentum dissipation rate, and hence to the DC resistivity, which is proportional to the entropy of the state.

This hydrodynamic description of a metallic state is far removed from the standard Fermi liquid theory of metals. The quasiparticles of a Fermi liquid interact weakly and thus the state has a very high viscosity.\cite{PinesNozieres} This means it cannot reach the local thermal equilibrium required for hydrodynamics to be valid before interactions with the lattice cause the state to lose momentum. We discuss in detail some experimentally observable consequences of this hydrodynamic mechanism and discuss its potential applicability to real systems, in particular the strange metallic phase of the cuprates.

For the reader uncomfortable with our assertion that a strongly interacting quantum critical system could behave like a hydrodynamic liquid with a minimal viscosity, we provide a proof of principle by discussing a specific classical gravitational theory, which arises as a low energy limit of string theory, that is dual to a field theory that exhibits this mechanism. This theory is locally quantum critical in the IR, i.e. it has a dynamical critical exponent $z\rightarrow\infty$. Analogous holographic theories controlled by a finite $z$ fixed point do not exhibit this mechanism, for reasons we will discuss.\cite{Lucas:2014zea} We also discuss how one can include the gross effects of the weak interaction of this holographic strange metal state with disorder by the inclusion of a graviton mass.\cite{Vegh:2013sk}

The remainder of the paper is structured as follows. In section \ref{sec:hydrocalculation}, we briefly summarise the memory matrix approach to transport in systems in which momentum is the only long-lived quantity and use this to calculate the momentum relaxation time, and hence the resistivity, in a hydrodynamic state weakly coupled to random disorder. We show that this has a viscous contribution which, in a strongly interacting quantum critical state, will be proportional to the entropy density of the state. The presence of this viscous contribution follows from a simple physical argument which can be understood without knowledge of the memory matrix formalism. In section \ref{sec:holographysection}, we then describe a holographic system in which, at sufficiently long distances, hydrodynamics is a good enough description that this mechanism is at work, as well as describing how one can directly include the main effects of quenched disorder in this setup. Finally, in section \ref{sec:discussionsection} we close by summarising our results and explaining their experimental implications, the potential applicability to real systems such as the strange metal phase of the cuprates, as well as some interesting directions for further work.

\section{Resistivity of a hydrodynamic fluid}
\label{sec:hydrocalculation}

As outlined in the introduction, we are interested in highly collective, charged states in which the intrinsic relaxation timescale of the electrical current is small. Despite these short relaxation timescales, the DC electrical resistivity in a translationally invariant theory of this kind will vanish. This is because the current carries a non-zero momentum which cannot decay, as it is a conserved Noether charge of the system. If we allow the momentum to decay slowly, in comparison to the intrinsic decay rate of the electrical current, by breaking the translational invariance of the system -- for example via weak interactions with random impurities or a lattice -- the decay rate of the electrical current at late times, and hence the DC resistivity of the state $\rho_{DC}$, will be proportional to the rate at which momentum dissipates $\Gamma_k$. In the following, we will express this dissipation rate in terms of the characteristic timescale $\tau_k$ over which the momentum decays: $\Gamma\equiv\tau_k^{-1}$. Furthermore, assuming that the interactions which dissipate momentum are weak (such that momentum lives for a long time), then at leading order in a perturbative expansion, $\tau_k^{-1}$ is determined by properties of the translationally invariant state. 

In a hydrodynamic state in which translational invariance is weakly broken, $\tau_k^{-1}$ depends on the viscosity $\eta$ of the state such that the DC resistivity will have a viscous contribution $\rho_{DC}(T)\sim\eta(T)$. Before we show this explicitly using the memory matrix formalism, we outline a simple physical argument which illustrates why this is the case. In a (relativistic) hydrodynamic liquid, momentum diffuses with a characteristic diffusion constant $D=\eta/\left(\epsilon+P\right)$, where $\epsilon$ and $P$ are the energy density and pressure of the state. In the non-relativistic limit, $\epsilon+P=m_en_e$ where $m_e$ and $n_e$ are the electron mass and number density respectively. \cite{Forster} If translational invariance is broken over a characteristic length scale $l$, then the characteristic timescale $\tau_k$ over which this diffusing momentum relaxes will be (see also references \onlinecite{2005JPhy4.131..255S,PhysRevLett.106.256804})
\begin{equation}
\frac{1}{\tau_k} = \frac{D}{l^2} = \frac{\eta}{\left(\epsilon+P\right)l^2}.
\label{monrelaxtime}
\end{equation}
This mechanism produces a DC resistivity $\rho_{DC}=\left(\epsilon+P\right)/\sigma^2\tau_k$, where $\sigma$ is the charge density of the state, or in the non-relativistic case $\rho_{DC}=1/\omega_p^2\tau_{k}$, where $\omega_p$ is the plasma frequency of the electrons.

This reasoning is valid for any hydrodynamic liquid, but we now specialise to the case of strongly interacting quantum critical liquids. These have the special property that there is a simple relationship between the shear viscosity and the entropy density $\eta/s=A\hbar/k_B$, where $A$ is a number of order one \cite{Sachdev} that in holographic theories is generically equal to $1/4\pi$. \cite{Kovtun:2004de,Iqbal:2008by} This minimal viscosity ensures that relaxation to local thermal equilibrium takes the minimal amount of time, which may be significantly shorter than the timescale for momentum loss due to Umklapp scattering or interactions with impurities. This is drastically different from an ordinary metal, in which fast Umklapp scattering of quasiparticles prevents the formation of a hydrodynamic state. \cite{PhysRevB.7.2317}

Combining these simple results, we arrive at a stunning conclusion:
\begin{equation}
\rho_{DC} (T) = \frac{A\hbar}{\omega_p^2m_el^2} \frac{S_e(T)}{k_B}.
\label{rhovsent}
\end{equation}
Having assumed only that the hydrodynamic limit is approximately valid and that the viscosity is minimal, we have found that the DC resistivity $\rho_{DC}$ due to this mechanism is proportional to the entropy per electron $S_e$.

\subsection{Memory matrix derivation}

We will now use a more rigorous memory matrix calculation to confirm our intuition that this effect should be present. Recent overviews of this formalism can be found in references \onlinecite{Hartnoll:2007ih,Hartnoll:2012rj,Mahajan:2013cja}. If the momentum of a system is its only long-lived quantity, and there is overlap between the electrical current and momentum operators in that theory, the DC electrical resistivity is related to the momentum relaxation time $\tau_k$ via\cite{Hartnoll:2012rj}
\begin{equation}
\rho_{DC}=\frac{\chi_{\vec{P}\vec{P}}}{\chi_{\vec{J}\vec{P}}^2}\;\tau_k^{-1}
\end{equation}
where $\chi_{\mathcal{O}_A\mathcal{O}_B}$ are the static susceptibilities of the state. The momentum relaxation time $\tau_k$ is determined by the microscopic processes which cause momentum to dissipate. Suppose we perturb a translationally invariant metal by the introduction of a spatially random potential $V\left(\vec{x}\right)$ for an operator $\mathcal{O}$ in the IR Hamiltonian (for definiteness, we have chosen here a (2+1)-dimensional metallic state)
\begin{equation}
\delta H=\int d^2\vec{x}\;V\left(\vec{x}\right)\mathcal{O}\left(t,\vec{x}\right)
\end{equation} 
where, on statistically averaging over the random potential,
\begin{equation}
\left<V\left(\vec{x}\right)\right>=0, \;\;\;\;\;\;\; \left<V\left(\vec{x}\right)V\left(\vec{y}\right)\right>=\bar{V}^2\delta^{(2)}\left(\vec{x}-\vec{y}\right).
\end{equation}
Such sources would naturally arise in the IR theory if there are random impurities present. Provided that the interaction with disorder is weak -- automatically true if the operator $\mathcal{O}$ is irrelevant in the IR -- this will cause momentum to dissipate slowly (so that it is still a long-lived quantity) and we can compute this dissipation rate perturbatively. At leading order, it is determined by the spectral weight of $\mathcal{O}$ at low energies \textit{in the translationally invariant state}\cite{Hartnoll:2007ih,Hartnoll:2008hs}
\begin{equation}
\label{eq:generalmemorymatrixresultfortau}
\tau_k^{-1}=\frac{\bar{V}^2}{2\chi_{\vec{P}\vec{P}}}\int \frac{d^2k}{\left(2\pi\right)^2}\;k^2 \lim_{\omega\rightarrow0}\frac{\text{Im}G_{\mathcal{O}\mathcal{O}}\left(\omega,k\right)}{\omega}\Biggr|_{\bar{V}\rightarrow0}
\end{equation}
where $G_{\mathcal{O}_A\mathcal{O}_B}$ denotes the retarded Greens function of two operators. This result has a simple physical interpretation: if translational invariance is weakly broken at a characteristic momentum scale $k$ by a source for $\mathcal{O}$, then at leading order in a perturbative expansion the rate at which momentum dissipates due to this should depend on the number of low energy excitations of the translationally invariant state with momentum $k$ that overlap with the operator $\mathcal{O}$, which is what the spectral weight tells us. The result above depends on the integral of the spectral weight over $k$, as the spatially random potential breaks translational invariance at all length scales.

It is expected that multiple operators may acquire a spatially dependent source in the low energy theory describing a real metal, in the presence of impurities and a lattice. In this situation, the argument above applies provided that all such operators couple weakly, and the most relevant of these will then provide the leading contribution to $\rho_{DC}$.

In this scenario, the microscopic quantities controlling the DC resistivity are the low energy spectral weights of the relevant operators in a given theory. Suppose that the low energy physics of the translationally invariant state is well-described by the laws of hydrodynamics. The important quantities in a (relativistic) hydrodynamic theory are the conserved energy-momentum tensor $T^{\mu\nu}$ and charge current $J^\mu$ of the theory. The form of the retarded Greens functions of these operators are specified by the laws of hydrodynamics (see, for example, reference \onlinecite{Kovtun:2012rj}), and are written in terms of the transport coefficients of the theory, such as the viscosity $\eta$, whose values are dependent upon the microscopic physics underlying the effective hydrodynamic description.

The presence of random impurities will give rise to a random source of energy density $T^{tt}$ in the low energy theory. If the coupling to random impurities is weak such that momentum dissipates slowly, this will produce a DC resistivity
\begin{equation}
\label{eq:neutralmemorymatrixrho}
\begin{aligned}
\rho_{DC}=\frac{\chi_{\vec{P}\vec{P}}}{\chi_{\vec{J}\vec{P}}^2}\;\tau_k^{-1}&=\frac{\bar{V}^2_{T^{tt}}}{2\chi_{\vec{J}\vec{P}}^2}\int \frac{d^2k}{\left(2\pi\right)^2}k^2\lim_{\omega\rightarrow0}\frac{\text{Im}G_{T^{tt}T^{tt}}\left(\omega,k\right)}{\omega}\Biggr|_{\bar{V}_{T^{tt}}\rightarrow0}\sim\frac{\bar{V}^2_{T^{tt}}}{\sigma^2}\int dk k\left(\eta k^2+\ldots\right),
\end{aligned}
\end{equation} 
where we have used the known spectral weight of $T^{tt}$ in a relativistic hydrodynamic theory with a conformally invariant vacuum\cite{Kovtun:2012rj} i.e. the stress-energy tensor is traceless (the pressure is related to the energy density by $\epsilon=2P$, and the bulk viscosity vanishes). This case is relevant for the holographic theories we will discuss shortly. In the final step of equation (\ref{eq:neutralmemorymatrixrho}) we have neglected order 1 numerical prefactors. The leading term in the small $k$ expansion on the right hand side is precisely the viscous contribution identified in equation (\ref{monrelaxtime}), but now derived in a rigorous manner. For a hydrodynamic theory with a minimal viscosity, this produces a DC resistivity $\rho_{DC}\left(T\right)\sim s\left(T\right)$. We are assuming that hydrodynamics is a good effective description down to a temperature-independent microscopic length scale, which acts as a UV cutoff on the momentum integrals, resulting in a controlled hydrodynamic momentum expansion.

If the random impurities lead to a significant source of charge density $J^t$ in the low energy theory, there will be additional contributions to the DC resistivity, taking the form
\begin{equation}
\label{eq:chargedmemorymatrixrho}
\rho_{DC}\sim\frac{\bar{V}^2_{J^{t}}}{\sigma^2}\int dkk\Bigl(\frac{1}{\sigma_Q}\left[2\frac{\sigma^2}{\epsilon+P}-\left(\frac{d\sigma}{d\mu}\right)_T\right]^2+k^2\frac{\sigma^2}{\left(\epsilon+P\right)^2}\eta+\ldots\Bigr),
\end{equation}
where we have again used the spectral weight of $J^t$ in a relativistic hydrodynamic theory with a conformally invariant vacuum.\cite{Kovtun:2012rj} There is also a viscous contribution in this case, in addition to a contribution inversely proportional to the ``universal conductivity'' transport coefficient $\sigma_Q$. Despite its name, the temperature dependence of $\sigma_Q$ is not universal. This illustrates an important point: there are various contributions to the DC resistivity of a hydrodynamic liquid, which will depend upon the microscopic details of the specific theory under consideration. We have highlighted the role of the viscous term because there are good theoretical and experimental reasons to expect it to make a universal contribution $\rho_{DC}\left(T\right)\sim\eta\left(T\right)\sim S\left(T\right)$, independent of such details, in strongly interacting, quantum critical systems. Additional contributions to equations (\ref{eq:neutralmemorymatrixrho}) and (\ref{eq:chargedmemorymatrixrho}) will also arise if one relaxes the requirement of a conformally invariant vacuum, includes higher order terms in the constitutive relations, or includes non-analytic effects.{\cite{Balasubramanian:2013yqa,Kovtun:2011np} Furthermore, if there are inhomogeneities sourced by a more relevant operator in the hydrodynamic theory than the energy density, their effects will be more important than those discussed here. These should be evaluated on a case-by-case basis.

We have shown rigorously, using the memory matrix formalism, that there should be a viscous contribution to the DC resistivity $\rho_{DC}\left(T\right)\sim\eta(T)$ when a hydrodynamic state is weakly coupled to disorder. Results of this nature have been derived previously by other methods.\cite{2005JPhy4.131..255S,PhysRevLett.106.256804} We note in particular reference \onlinecite{PhysRevLett.106.256804}, in which an expression similar to equation (\ref{eq:chargedmemorymatrixrho}) was found, but with the leading term inversely proportional to the thermal conductivity. This is consistent, as the thermal conductivity is proportional to the universal conductivity $\sigma_Q$ in the type of theories we have considered here.\cite{Herzog:2009xv} In the cuprates, it is realistic that this term is small (in comparison to the contribution from neutral disorder in equation (\ref{eq:neutralmemorymatrixrho})) due to the chemistry of these systems.\cite{PhysRevLett.106.256804,RevModPhys.81.45}

\section{Holographic realisation of this mechanism}
\label{sec:holographysection}

We have shown that if the low energy physics of a system is well-described by hydrodynamics with a minimal viscosity $\eta\sim s$, there will be a viscous contribution to the DC resistivity such that $\rho_{DC}\sim\eta\sim s$, when this theory is weakly coupled to disorder. To motivate the applicability of these results to strongly correlated electron systems and, in particular, the strange metal phase of the cuprates, we will now describe a well-controlled example of a strongly interacting, quantum critical state of matter with these features. We give this example as a proof of principle, although we do not expect that its specific properties -- in particular, that it can be described by a dual, classical theory of gravity -- are necessary for the existence of the effective hydrodynamic description which is required. For the reader unfamiliar with using holographic techniques to study strongly interacting states, the references \onlinecite{McGreevy:2009xe,Hartnoll:2009sz,Herzog:2009xv,Hartnoll:2011fn} provide informative introductions to this topic.

\subsection{Translationally invariant theory}

As indicated above, the example we give of a strongly interacting, quantum critical state with these properties is equivalent, by a holographic duality, to a classical theory of gravity in a curved, higher dimensional spacetime. This (2+1)-dimensional `strange metal' is a finite density state which exhibits local quantum criticality at low energies. Local quantum criticality is an emergent quantum temporal scale invariance but with short-ranged spatial correlations, and is closely related to the marginal Fermi liquid phenomenology proposed to explain experimental observations in the cuprates.\cite{PhysRevLett.63.1996} States with this property are common in holography, and they typically exhibit superconducting instabilities,\cite{Gubser:2008px,Hartnoll:2008vx} non-Fermi-liquid behaviour of fermionic two-point functions,\cite{Liu:2009dm,Cubrovic:2009ye,Faulkner:2009wj,Gubser:2009qt} and interesting power laws in the mid-IR optical conductivity in the presence of a lattice.\cite{Horowitz:2012ky,Horowitz:2012gs,Ling:2013nxa}

This state is dual to a black brane solution of a (3+1)-dimensional Einstein-Maxwell-Dilaton theory of classical gravity that was first studied in this context by Gubser and Rocha,\cite{Gubser:2009qt} and which can be uplifted to a solution of 11-dimensional supergravity \cite{Cvetic:1999xp}
\begin{equation}
S_{EMD}=\frac{1}{2\kappa_4^2}\int d^4x\sqrt{-g}\left[\mathcal{R}-\frac{1}{4}e^\phi F_{\mu\nu}F^{\mu\nu}-\frac{3}{2}\partial_\mu\phi\partial^\mu\phi+\frac{6}{L^2}\cosh\phi\right].
\end{equation}
The charged black brane solution of this theory is
\begin{equation}
\label{eq:masslessholographicsolution}
\begin{aligned}
ds^2&=\frac{r^2g(r)}{L^2}\left(-h(r)dt^2+dx^2+dy^2\right)+\frac{L^2}{r^2g(r)h(r)}dr^2, \;\;\;\;\; h(r)=1-\frac{\left(Q+r_0\right)^3}{\left(Q+r\right)^3},\\
A_t(r)&=\sqrt{\frac{3Q\left(Q+r_0\right)}{L^2}}\left(1-\frac{Q+r_0}{Q+r}\right), \;\;\;\;\;\phi(r)=\frac{1}{3}\log(g(r)),\;\;\;\;\;g(r)=\left(1+\frac{Q}{r}\right)^\frac{3}{2}.
\end{aligned}
\end{equation}
As usual, the ``holographic'' co-ordinate $r$ labels the additional dimension in which the gravitational theory is defined compared to the field theory, and $r_0$ denotes the location of the black brane horizon.

Note that in addition to the minimal ingredients required for a holographic dual of a quantum field theory at non-zero density (the metric $g_{\mu\nu}$ and a U(1) gauge field $A_{\mu}$), we include a scalar field $\phi$ (the dilaton) which means that the field theory contains a neutral scalar operator with non-trivial dynamics. Although the gravitational theory is simpler in the absence of the dilaton, the price of this simplicity is that it is dual to a field theory state which is unrealistic: it has a finite zero temperature entropy. The presence of the dilaton changes the ground state, which now has vanishing zero temperature entropy. At finite temperature, it has an entropy density which is linear in temperature $s\sim T$ at low $T$. We note here that there are many other locally quantum critical, holographic examples of this mechanism which produces $\rho_{DC}\sim s$ -- we have chosen to describe a case in which $s\sim T$ as this is the case relevant to the experimental system we are addressing: the strange metal phase of the cuprates.

The low energy, local quantum criticality of this state is encoded in the near-horizon geometry of the dual black brane solution at zero temperature.\cite{Iqbal:2011in} This geometry is conformal to AdS$_2\times\mathbb{R}^2$ and therefore transforms covariantly (the line element transforms as $ds^2\rightarrow\lambda^{-1}ds^2$) under a scaling symmetry which acts on the temporal co-ordinate $t$, but not on the spatial co-ordinates $(x,y)$, of the strongly interacting state
\begin{equation}
\label{eq:scalingsymmetries}
t\rightarrow\lambda t,\;\;\;\;\;\;x\rightarrow x,\;\;\;\;\;\;y\rightarrow y,\;\;\;\;\;\;r\rightarrow \lambda^{-2}r.
\end{equation}
This state violates hyperscaling and in the standard classification in terms of a dynamical critical exponent $z$ and hyperscaling violation exponent $\theta$, this state has $z\rightarrow\infty$ with the ratio $-\theta/z=1$ fixed.\cite{Goldstein:2009cv,Charmousis:2010zz,Hartnoll:2012wm}

\subsection{Weak interactions with random disorder}

The low energy spectral functions of operators in the thermal state of this theory are of the form\cite{Hartnoll:2012rj,Anantua:2012nj}
\begin{equation}
\label{eq:holographicgreensfunctionmatching}
\lim_{\omega\rightarrow0}\text{Im}G_{\mathcal{O}\mathcal{O}}\left(\omega,k,T\right)\sim H(k)\lim_{\omega\rightarrow0}\text{Im}\mathcal{G}^{IR}_{\mathcal{O}\mathcal{O}}\left(\omega,k,T\right)\sim H(k)\;\omega\;T^{2\nu_\mathcal{O}\left(k\right)-1}.
\end{equation}
At low $\omega$ and $T$, the $\omega$ and $T$ dependence of the spectral function is contained in the ``IR spectral function'' $\mathcal{G}^{IR}\left(\omega,k,T\right)$, whose form is determined by the near-horizon geometry in the dual gravitational theory. The IR local quantum criticality of this state means that this IR spectral function is a power law in $\omega$ and $T$, and is approximately momentum-independent. The only momentum dependence is in the exponent $\nu_\mathcal{O}\left(k\right)$, which is determined by the mass, in the near-horizon geometry, of the excitation of the field dual to the operator $\mathcal{O}$. To determine the low energy spectral function of an operator in the field theory, one must additionally determine the momentum-dependent ``matching function'' $H(k)$ by matching the solutions of the field equations for excitations in the near-horizon geometry to the asymptotic region of the geometry.

The advantage of the decomposition of the Greens function given in equation (\ref{eq:holographicgreensfunctionmatching}) is that, to determine the leading $T$-dependence of the spectral function of $\mathcal{O}$, the matching to the asymptotic region does not need to be explicitly performed. One must only determine the exponent $\nu_\mathcal{O}\left(k\right)$, which is related to the scaling dimension of $\mathcal{O}$ at the locally critical IR fixed point and is fixed by the near-horizon properties of the dual gravitational theory. This property can be used to determine the DC resistivity of these states,\cite{Hartnoll:2012rj,Anantua:2012nj} as we will now review.

The translationally invariant state dual to the solution in equation (\ref{eq:masslessholographicsolution}) has a vanishing DC resistivity due to the conservation of momentum. If spatially random sources are added for $\mathcal{O}$, they cause momentum to dissipate at a rate
\begin{equation}
\Gamma_k\equiv\tau_k^{-1}\sim\frac{\bar{V}_\mathcal{O}^2}{\chi_{\vec{P}\vec{P}}}\int dk\;k^3\; H(k)\; T^{2\nu_\mathcal{O}\left(k\right)-1},
\end{equation}
provided that $\mathcal{O}$ is weakly coupled at the low energies and temperatures of interest and we can therefore use equation (\ref{eq:generalmemorymatrixresultfortau}). The introduction of random impurities in a sample will induce a spatially random source of energy density $T^{tt}$, which therefore produces a DC resistivity
\begin{equation}
\rho_{DC}=\frac{\chi_{\vec{P}\vec{P}}}{\chi_{\vec{J}\vec{P}}^2}\tau_k^{-1}\sim\frac{\bar{V}_{T^{tt}}^2}{\sigma^2}\int dk\;k^3\;H(k)\;T^{2\nu_{T^{tt}}\left(k\right)-1},
\end{equation} 
where the appropriate exponent for this operator is 
\begin{equation}
\nu_{T^{tt}}\left(k\right)\equiv\nu\left(k\right)=\sqrt{\frac{11}{3}+4\frac{k^2}{\mu^2}-\frac{8}{3}\sqrt{1+3\frac{k^2}{\mu^2}}},
\end{equation}
where $\mu=\sqrt{3}Q/L^2$ is the chemical potential of the state. At leading order in $T$, the dominant contribution to this integral comes from the homogeneous $k=0$ mode of the disorder such that
\begin{equation}
\label{eq:linearresistivityoperatordimensionresult}
\rho_{DC}\left(T\right)\sim T^{2\nu\left(0\right)-1}\sim T.
\end{equation}
Including the weak momentum dependence of $\nu\left(k\right)$ in the integral produces small logarithmic corrections to this leading order result such that $\rho_{DC}\sim T\log\left(\mu/T\right)^{-1}$. For more general locally critical holographic states, it is implicitly contained in the results of reference \onlinecite{Anantua:2012nj} that this calculation yields $\rho_{DC}\left(T\right)\sim s\left(T\right)$ at leading order. The operators $J^t$ and $T^{tt}$ have the same exponent $\nu\left(k\right)$ as their dual fields are coupled in the IR geometry, and thus a spatially random source for charge density will also produce a contribution $\rho_{DC}\left(T\right)\sim s\left(T\right)$. Finally, we note that both of these operators are (marginally) irrelevant at low energies and thus this approach is consistent.

When analysed in this way, it appears that the linear resistivity of this state is highly dependent upon the microscopic details of the theory, in particular the dimensions of operators in the low energy theory, which are calculated from masses of field excitations in a higher-dimensional gravitational theory. It is not obvious whether these features could realistically be expected in real electron systems, such as in the strange metal phase of the cuprates, or whether they arise only in the special class of strongly interacting field theories which have dual gravitational descriptions.

We can get another perspective on this result by revisiting the spectral functions of these holographic locally critical quantum field theories. By explicitly calculating the ``matching functions'' $H(k)$ numerically, one finds that the Greens functions of these theories are consistent with those of hydrodynamics, with a minimal viscosity, down to length scales $k\lesssim\mu$.\cite{Davison:2013bxa,Davison:2013uha,Tarrio:2013tta} The most explicit verification of this is in the simplest example of the Reissner-Nordstrom-AdS$_4$ solution of Einstein-Maxwell theory where one can analytically compute the matching functions $H(k)$ and determine that hydrodynamics is a good description of this theory, in the limit where the finite $k$ corrections to $\nu\left(k\right)$ can be neglected.\cite{Davison:2013bxa} We expect this to be true more generally in holographic locally critical states, and it would be interesting to explicitly verify this. There is therefore a contribution to the DC resistivity $\rho_{DC}\left(T\right)\sim s\left(T\right)$ via the viscous mechanism described in section \ref{sec:hydrocalculation}. This gives a more physical understanding of the result in equation (\ref{eq:linearresistivityoperatordimensionresult}). Fundamentally, if the state is described by hydrodynamics then the scaling dimensions of $T^{tt}$, $J^t$ etc., which were used to compute $\rho_{DC}$ in reference \onlinecite{Anantua:2012nj}, are not arbitrary numbers but are constrained by the transport coefficients (such as the viscosity) of the hydrodynamic description.

The finite $k$ corrections to $\nu\left(k\right)$ in these theories, which are not captured by hydrodynamics, produce small, logarithmic corrections to this result as previously explained. These corrections reflect the fact that at low energies and finite momenta, this holographic state has slightly \textit{less} spectral weight than relativistic hydrodynamics (since $\nu_k\ge\nu_0$) and it would be interesting to search for such corrections experimentally via precision measurements of $\rho_{DC}$ over a large range of $T$. We note that coupling a hydrodynamic liquid to a periodic potential will also give a viscous contribution to the DC resistivity $\rho_{DC}\sim s$ with $l$ in equation (\ref{rhovsent}) given by the lattice spacing. In the state dual to the gravitational solution (\ref{eq:masslessholographicsolution}), one instead finds that the DC resistivity $\rho_{DC}\sim T^{2\nu\left(k_L\right)-1}$ obeys a power law where the power depends upon $k_L$.\cite{Anantua:2012nj} In these circumstances, the non-hydrodynamic finite-$k$ corrections to $\nu\left(k\right)$ are important and thus an effective hydrodynamic description is not accurate enough. An exception to this is when $k_L\ll\mu$, in which case $\nu\left(k_L\right)\sim\nu\left(0\right)$, in which case $\rho_{DC}\sim T$ as was noted in a footnote in reference \onlinecite{Anantua:2012nj}.


Finally, we note that although the hydrodynamic underpinning should ensure the universality of our mechanism, its subsequent expression for the resistivity does rely on the additional assumption of local quantum criticality. For critical theories with a finite dynamical critical exponent $z$ -- where locally quantum critical means $z=\infty$ -- the UV-cut-off in the momentum integral in the memory matrix expressions (\ref{eq:neutralmemorymatrixrho}) and (\ref{eq:chargedmemorymatrixrho}) may become temperature dependent, see e.g. reference \onlinecite{Lucas:2014zea} for an example of this in a different set-up.

\subsection{Explicit inclusion of momentum dissipation}

As we have just reviewed, although random disorder breaks translational invariance on all length scales, the leading contribution to the momentum dissipation rate $\Gamma_k$ and hence to $\rho_{DC}$ in these holographic locally quantum critical states is from the homogeneous $k=0$ mode of this disorder. We can explicitly incorporate its effects on $\rho_{DC}$ in the holographic theory via the inclusion of a graviton mass term in the action\cite{Vegh:2013sk,Davison:2013jba,Blake:2013bqa}
\begin{equation}
\begin{aligned}
\label{eq:specificmassivegravityaction}
S=\frac{1}{2\kappa_4^2}\int d^4x\sqrt{-g}\Bigl[\mathcal{R}-\frac{1}{4}e^\phi F_{\mu\nu}F^{\mu\nu}-\frac{3}{2}\partial_\mu\phi\partial^\mu\phi+\frac{6}{L^2}\cosh\phi-\frac{1}{2}m^2\left(\text{Tr}\left(\mathcal{K}\right)^2-\text{Tr}\left(\mathcal{K}^2\right)\right)\Bigr],
\end{aligned}
\end{equation}
where $\mathcal{K}^\mu_\alpha\mathcal{K}^\alpha_\nu=g^{\mu\alpha}f_{\alpha\nu}$, and the non-zero elements of the fixed reference metric $f_{\mu\nu}$ are $f_{xx}=f_{yy}=1$. This explicit breaking of diffeomorphism invariance in the gravitational theory is equivalent to the loss of momentum conservation in the dual field theory. This theory has a planar black hole solution
\begin{equation}
\begin{aligned}
\label{eq:massivegravitysolution}
ds^2&=\frac{r^2g(r)}{L^2}\left(-h(r)dt^2+dx^2+dy^2\right)+\frac{L^2}{r^2g(r)h(r)}dr^2,\\
A_t(r)&=\sqrt{\frac{3Q\left(Q+r_0\right)}{L^2}\left(1-\frac{m^2L^4}{2\left(Q+r_0\right)^2}\right)}\left(1-\frac{Q+r_0}{Q+r}\right),\;\;\;\;\; \phi(r)=\frac{1}{3}\log(g(r)),\\
h(r)&=1-\frac{m^2L^4}{2\left(Q+r\right)^2}-\frac{\left(Q+r_0\right)^3}{\left(Q+r\right)^3}\left(1-\frac{m^2L^4}{2\left(Q+r_0\right)^2}\right),\;\;\;\;\;\;\;g(r)=\left(1+\frac{Q}{r}\right)^\frac{3}{2},
\end{aligned}
\end{equation}
where the radial co-ordinate $r$ spans the range between $r_0$, the location of the horizon, and $\infty$, the boundary of the spacetime. The temperature $T$ and chemical potential $\mu$ of the dual field theory are
\begin{equation}
\begin{aligned}
T=\frac{r_0\left(6\left(1+Q/r_0\right)^2-\frac{m^2L^4}{r_0^2}\right)}{8\pi L^2\left(1+Q/r_0\right)^{3/2}},\;\;\;\;\;\;\;\;\;\;\mu=\frac{\sqrt{3Q\left(Q+r_0\right)\left(1-\frac{m^2L^4}{2\left(Q+r_0\right)^2}\right)}}{L^2}.
\end{aligned}
\end{equation}
Even with $m\ne0$, the zero temperature near-horizon geometry of (\ref{eq:massivegravitysolution}) retains the scaling symmetries (\ref{eq:scalingsymmetries}).

For the chemical potential $\mu$ to be real, we require that $m^2L^4\le2\left(Q+r_0\right)^2$, and therefore $T/\mu\propto\sqrt{r_0/Q}$ at low $T/\mu$. The linear entropy density $s$ of the system at low $T$ can be made transparent by writing it as a function of $r_0/Q$ and $\bar{m}\equiv m/\mu$
\begin{equation}
\begin{aligned}
s/\mu^2=\frac{2\pi L^2}{3\kappa_4^2}\sqrt{r_0/Q}\sqrt{1+r_0/Q}\left(1+\frac{3\bar{m}^2}{2\left(1+r_0/Q\right)}\right)\sim T/\mu \text{ at low } T,
\end{aligned}
\end{equation}
and the charge density $\sigma$ is
\begin{equation}
\begin{aligned}
\sigma/\mu^2=\frac{L^2}{2\sqrt{3}\kappa_4^2}\sqrt{1+r_0/Q}\sqrt{1+\frac{3\bar{m}^2}{2\left(1+r_0/Q\right)}}\sim \left(T/\mu\right)^0 \text{ at low } T,
\end{aligned}
\end{equation}
where we have calculated these from the area of the horizon and Gauss' law respectively. 

The universal result of Blake and Tong \cite{Blake:2013bqa} then ensures a linear resistivity $\rho_{DC}$ at low temperatures
\begin{equation}
\label{eq:analyticmassivegravityresistance}
\rho_{DC}=\frac{s}{4\pi\sigma^2}m^2=\frac{2\kappa_4^2}{L^2}\frac{1}{\sqrt{1+Q/r_0}}\frac{m^2}{\mu^2}\sim T/\mu \text{ at low } T,
\end{equation}
as in the cuprate strange metal phase. The graviton mass produces a non-zero $\rho_{DC}$ by coupling the momentum to a uniform operator i.e. one with no characteristic periodic length scale\cite{Davison:2013jba} -- one consequence of this is that the planar black hole solution is homogeneous and isotropic. This operator plays the role of the homogeneous mode sourced by random disorder in $T^{tt}$ (or $J^t$), producing a DC resistivity linear in $T$. The arbitrary dimensionful parameter in the holographic setup (the graviton mass $m$) corresponds to the energy scale characterising the impurities i.e. $l^{-1}$ in equation (\ref{rhovsent}), or $\bar{V}$ in equation (\ref{eq:neutralmemorymatrixrho}). It must be small $m\ll\mu$ for the coupling to impurities to be weak in the IR.

There is a lot of evidence that massive theories of gravity arise generically as the effective description of the low energy physics in holographic systems in which momentum dissipates slowly. That is, translational symmetry breaking generates an effective mass for metric components which controls the DC resistivity of the state. This is certainly the case when translational symmetry is broken by sources for a scalar operator.\cite{Andrade:2013gsa,Blake:2013owa,Lucas:2014zea,Gouteraux:2014hca,Donos:2014uba}  One way to generate a graviton mass like that above\cite{Andrade:2013gsa} (rather than by inserting it by hand into the action) is to break translational invariance explicitly, in a homogeneous and isotropic way, via the introduction of linear sources for two additional scalar fields $\varphi^i=mx^i$. This effective theory does not capture the subleading effects of the non-homogeneous components of the disorder: see reference \onlinecite{Lucas:2014zea} for work in this direction. It would be interesting to try and generalise these results to the cases where translational invariance is explicitly broken in the UV not by scalars, but by the metric or gauge field. This would constitute an explicit proof that theories of massive gravity are always the relevant low energy effective theory for holographic systems with slow momentum dissipation.

\section{Discussion}
\label{sec:discussionsection}

In summary, we have presented a simple mechanism which can explain how strange metals with a linear in temperature electronic entropy can acquire a linear in temperature resistivity. It relies upon the assumption that the electronic system is well described by the laws of hydrodynamics with a minimal viscosity, a feature that arises naturally in systems with a holographic dual. While experimental tests will ultimately determine whether this explanation is correct, holography has once again proved a valuable tool for thinking about old problems in new ways. We outline below some of the implications of our mechanism and some experimental tests of it. As we have previously mentioned, a hydrodynamic description is not applicable at short distances in a conventional metal, and it would be very interesting to determine further ``smoking gun'' experimental signals of hydrodynamic behaviour in metals. We hope to return to this point in the future. 

The mechanism we have described offers remarkably simple explanations for some mysterious experimental facts in the high $T_c$ cuprate superconductors. Firstly, the linear resistivity of the strange metal phase is due to its linear electronic entropy.\cite{PhysRevLett.71.1740} Secondly, the vanishing residual resistivity (see, for example, reference \onlinecite{2013PNAS..11012235B}) is due to the vanishing entropy when $T=0$: in this limit $\eta$ vanishes and thus this perfect fluid does not dissipate momentum. Experimentally,\cite{vdmarelpowerlawcuprates} it is known that $1/\tau_k \simeq \hbar / k_B T$ and using equation (\ref{rhovsent}), we estimate that $l \simeq 10^{-9}$ m, remarkably close to the Ioffe-Regel limit ($l\simeq a$, the lattice constant). This sheds new light on the long-standing puzzle of why the strong chemical disorder which is known to be intrinsic to cuprate crystals is not imprinted on their residual resistivity. It may also explain why the effects of electron-phonon couplings appear to be invisible in the electronic transport properties. At non-zero temperatures, there should be a strongly temperature-dependent contribution to momentum relaxation due to inelastic scattering against phonons, but in a dirty metal this is overwhelmed by the temperature-independent elastic scattering. If we can incorporate both effects by an effective mean free path $l$ in equation (\ref{monrelaxtime}), the effects of the electron-phonon interactions will be invisible.

The mechanism we have described here may be tested experimentally. The $T^2$ resistivity of the cuprates in the pseudogap regime\cite{2013PNAS..11012235B,2013PNAS..110.5774M} coincides with a reduction in the electronic specific heat to a form which looks quadratic in temperature.\cite{PhysRevLett.71.1740} If the entropy law (\ref{rhovsent}) is controlling this resistivity, then systematic, precision measurements over a large range of $T$ should show a very close correlation between these quantities. The pseudogap regime is associated with ordering phenomena.\cite{2013Natur.498...75S} To have $\rho_{DC}\sim s$ in this regime, we require some novel, emergent, quantum critical degrees of freedom to be present. Although these are alien to the conventional theories of ordering, symmetry breaking is a ubiquitous phenomenon in holographic theories where these IR degrees of freedom abound. A second experimental signature of hydrodynamic behaviour in metals is a strong violation of the Wiedemann-Franz law.\cite{Mahajan:2013cja} To measure this would require suppressing $T_c$ with a large magnetic field in order to isolate the electronic contribution to the thermal conductivity. 

Finally, let us emphasise one of our key assumptions, and some questions that remain to be addressed. We argued that, unlike a Fermi liquid, a strongly interacting quantum critical system can form a hydrodynamic state before its interactions with a periodic potential become important, due to its minimal viscosity. However, given that the cuprates can form Mott insulators, the effects of a periodic potential are, a priori, expected to be large. For the holographic state studied here, the timescale $\tau^{}_U$ over which momentum is lost due to interactions with a periodic potential is $\tau_U^{-1}\sim T^{2\nu\left(k_L\right)-1}$ and this timescale may be significantly longer than the corresponding timescale due to quenched disorder in equation (\ref{monrelaxtime}).\cite{Hartnoll:2012rj}  This is required for our explanation to be valid.

Our mechanism cannot explain either the high value of $T_c$ or the $T$ dependence of the Hall angle \cite{PhysRevLett.67.2088,PhysRevLett.67.2092} in the cuprates. In particular, an explanation of the Hall angle may require the presence of an independent relaxation time associated to Hall transport, unlike in the hydrodynamic model we have outlined in which all transport is controlled by the momentum relaxation time. Furthermore, as highlighted in reference \onlinecite{2013Sci...339..804B}, linear resistivities with a ``Planckian"  momentum relaxation rate occur in a large variety of systems, including simple metals in the phonon-dominated regime, and heavy fermion-like systems. Phonon domination is the most deadly adversary of the hydrodynamic liquid we have described, while the heavy fermion systems acquire their name from their large specific heats that tend to diverge upon decreasing the temperature towards their quantum critical points. Moreover, these systems are much cleaner than the cuprates and we therefore do not see any reason to believe that our mechanism applies in these cases.

\begin{acknowledgments}
We are grateful to Mike Blake, Chris Herzog, Andrei Parnachev, Giuseppe Policastro, David Tong, David Vegh and especially Sean Hartnoll and Subir Sachdev for helpful discussions, and to the Isaac Newton Institute, Cambridge for hospitality while this work was completed. This research was supported in part  by a VIDI grant and a VICI grant from the Netherlands Organization for Scientific Research (NWO), by the Netherlands Organization for Scientific Research/Ministry of Science and Education (NWO/OCW) and by the Foundation for Research into Fundamental Matter (FOM). This publication was made possible through the support of a grant from the John Templeton Foundation. The opinions expressed in this publication are those of the authors and do not necessarily reflect the views of the John Templeton Foundation.
\end{acknowledgments}

\bibliographystyle{apsrev}
\bibliography{SMHydroPRB2}

\begin{thebibliography}{63}
\expandafter\ifx\csname natexlab\endcsname\relax\def\natexlab#1{#1}\fi
\expandafter\ifx\csname bibnamefont\endcsname\relax
  \def\bibnamefont#1{#1}\fi
\expandafter\ifx\csname bibfnamefont\endcsname\relax
  \def\bibfnamefont#1{#1}\fi
\expandafter\ifx\csname citenamefont\endcsname\relax
  \def\citenamefont#1{#1}\fi
\expandafter\ifx\csname url\endcsname\relax
  \def\url#1{\texttt{#1}}\fi
\expandafter\ifx\csname urlprefix\endcsname\relax\def\urlprefix{URL }\fi
\providecommand{\bibinfo}[2]{#2}
\providecommand{\eprint}[2][]{\url{#2}}

\bibitem[{\citenamefont{Zaanen}(2010)}]{Zaanen:2010yk}
\bibinfo{author}{\bibfnamefont{J.}~\bibnamefont{Zaanen}}
  (\bibinfo{year}{2010}), \eprint{1012.5461}.

\bibitem[{\citenamefont{Varma et~al.}(1989)\citenamefont{Varma, Littlewood,
  Schmitt-Rink, Abrahams, and Ruckenstein}}]{PhysRevLett.63.1996}
\bibinfo{author}{\bibfnamefont{C.~M.} \bibnamefont{Varma}},
  \bibinfo{author}{\bibfnamefont{P.~B.} \bibnamefont{Littlewood}},
  \bibinfo{author}{\bibfnamefont{S.}~\bibnamefont{Schmitt-Rink}},
  \bibinfo{author}{\bibfnamefont{E.}~\bibnamefont{Abrahams}}, \bibnamefont{and}
  \bibinfo{author}{\bibfnamefont{A.~E.} \bibnamefont{Ruckenstein}},
  \bibinfo{journal}{Phys. Rev. Lett.} \textbf{\bibinfo{volume}{63}},
  \bibinfo{pages}{1996} (\bibinfo{year}{1989}).

\bibitem[{\citenamefont{Maldacena}(1998)}]{Maldacena:1997re}
\bibinfo{author}{\bibfnamefont{J.~M.} \bibnamefont{Maldacena}},
  \bibinfo{journal}{Adv.Theor.Math.Phys.} \textbf{\bibinfo{volume}{2}},
  \bibinfo{pages}{231} (\bibinfo{year}{1998}), \eprint{hep-th/9711200}.

\bibitem[{\citenamefont{Faulkner et~al.}(2010)\citenamefont{Faulkner, Iqbal,
  Liu, McGreevy, and Vegh}}]{Faulkner:2010zz}
\bibinfo{author}{\bibfnamefont{T.}~\bibnamefont{Faulkner}},
  \bibinfo{author}{\bibfnamefont{N.}~\bibnamefont{Iqbal}},
  \bibinfo{author}{\bibfnamefont{H.}~\bibnamefont{Liu}},
  \bibinfo{author}{\bibfnamefont{J.}~\bibnamefont{McGreevy}}, \bibnamefont{and}
  \bibinfo{author}{\bibfnamefont{D.}~\bibnamefont{Vegh}},
  \bibinfo{journal}{Science} \textbf{\bibinfo{volume}{329}},
  \bibinfo{pages}{1043} (\bibinfo{year}{2010}).

\bibitem[{\citenamefont{Donos and Hartnoll}(2012)}]{Donos:2012ra}
\bibinfo{author}{\bibfnamefont{A.}~\bibnamefont{Donos}} \bibnamefont{and}
  \bibinfo{author}{\bibfnamefont{S.~A.} \bibnamefont{Hartnoll}},
  \bibinfo{journal}{Phys.Rev.} \textbf{\bibinfo{volume}{D86}},
  \bibinfo{pages}{124046} (\bibinfo{year}{2012}), \eprint{1208.4102}.

\bibitem[{\citenamefont{Hoyos et~al.}(2013)\citenamefont{Hoyos, Kim, and
  Oz}}]{Hoyos:2013eza}
\bibinfo{author}{\bibfnamefont{C.}~\bibnamefont{Hoyos}},
  \bibinfo{author}{\bibfnamefont{B.~S.} \bibnamefont{Kim}}, \bibnamefont{and}
  \bibinfo{author}{\bibfnamefont{Y.}~\bibnamefont{Oz}}, \bibinfo{journal}{JHEP}
  \textbf{\bibinfo{volume}{1311}}, \bibinfo{pages}{145} (\bibinfo{year}{2013}),
  \eprint{1304.7481}.

\bibitem[{\citenamefont{Hartnoll et~al.}(2007)\citenamefont{Hartnoll, Kovtun,
  Muller, and Sachdev}}]{Hartnoll:2007ih}
\bibinfo{author}{\bibfnamefont{S.~A.} \bibnamefont{Hartnoll}},
  \bibinfo{author}{\bibfnamefont{P.~K.} \bibnamefont{Kovtun}},
  \bibinfo{author}{\bibfnamefont{M.}~\bibnamefont{Muller}}, \bibnamefont{and}
  \bibinfo{author}{\bibfnamefont{S.}~\bibnamefont{Sachdev}},
  \bibinfo{journal}{Phys.Rev.} \textbf{\bibinfo{volume}{B76}},
  \bibinfo{pages}{144502} (\bibinfo{year}{2007}), \eprint{0706.3215}.

\bibitem[{\citenamefont{Hartnoll and Herzog}(2008)}]{Hartnoll:2008hs}
\bibinfo{author}{\bibfnamefont{S.~A.} \bibnamefont{Hartnoll}} \bibnamefont{and}
  \bibinfo{author}{\bibfnamefont{C.~P.} \bibnamefont{Herzog}},
  \bibinfo{journal}{Phys.Rev.} \textbf{\bibinfo{volume}{D77}},
  \bibinfo{pages}{106009} (\bibinfo{year}{2008}), \eprint{0801.1693}.

\bibitem[{\citenamefont{Hartnoll and Hofman}(2012)}]{Hartnoll:2012rj}
\bibinfo{author}{\bibfnamefont{S.~A.} \bibnamefont{Hartnoll}} \bibnamefont{and}
  \bibinfo{author}{\bibfnamefont{D.~M.} \bibnamefont{Hofman}},
  \bibinfo{journal}{Phys.Rev.Lett.} \textbf{\bibinfo{volume}{108}},
  \bibinfo{pages}{241601} (\bibinfo{year}{2012}), \eprint{1201.3917}.

\bibitem[{\citenamefont{Anantua et~al.}(2013)\citenamefont{Anantua, Hartnoll,
  Martin, and Ramirez}}]{Anantua:2012nj}
\bibinfo{author}{\bibfnamefont{R.~J.} \bibnamefont{Anantua}},
  \bibinfo{author}{\bibfnamefont{S.~A.} \bibnamefont{Hartnoll}},
  \bibinfo{author}{\bibfnamefont{V.~L.} \bibnamefont{Martin}},
  \bibnamefont{and} \bibinfo{author}{\bibfnamefont{D.~M.}
  \bibnamefont{Ramirez}}, \bibinfo{journal}{JHEP}
  \textbf{\bibinfo{volume}{1303}}, \bibinfo{pages}{104} (\bibinfo{year}{2013}),
  \eprint{1210.1590}.

\bibitem[{\citenamefont{Mahajan et~al.}(2013)\citenamefont{Mahajan, Barkeshli,
  and Hartnoll}}]{Mahajan:2013cja}
\bibinfo{author}{\bibfnamefont{R.}~\bibnamefont{Mahajan}},
  \bibinfo{author}{\bibfnamefont{M.}~\bibnamefont{Barkeshli}},
  \bibnamefont{and} \bibinfo{author}{\bibfnamefont{S.~A.}
  \bibnamefont{Hartnoll}}, \bibinfo{journal}{Phys.Rev.}
  \textbf{\bibinfo{volume}{B88}}, \bibinfo{pages}{125107}
  (\bibinfo{year}{2013}), \eprint{1304.4249}.

\bibitem[{\citenamefont{Hartnoll et~al.}(2014)\citenamefont{Hartnoll, Mahajan,
  Punk, and Sachdev}}]{Hartnoll:2014gba}
\bibinfo{author}{\bibfnamefont{S.~A.} \bibnamefont{Hartnoll}},
  \bibinfo{author}{\bibfnamefont{R.}~\bibnamefont{Mahajan}},
  \bibinfo{author}{\bibfnamefont{M.}~\bibnamefont{Punk}}, \bibnamefont{and}
  \bibinfo{author}{\bibfnamefont{S.}~\bibnamefont{Sachdev}},
  \bibinfo{journal}{Phys.Rev.} \textbf{\bibinfo{volume}{B89}},
  \bibinfo{pages}{155130} (\bibinfo{year}{2014}), \eprint{1401.7012}.

\bibitem[{\citenamefont{Sachdev}(2011)}]{Sachdev}
\bibinfo{author}{\bibfnamefont{S.}~\bibnamefont{Sachdev}},
  \emph{\bibinfo{title}{{Quantum Phase Transitions}}}
  (\bibinfo{publisher}{Cambridge University Press}, \bibinfo{year}{2011}).

\bibitem[{\citenamefont{Kovtun et~al.}(2005)\citenamefont{Kovtun, Son, and
  Starinets}}]{Kovtun:2004de}
\bibinfo{author}{\bibfnamefont{P.}~\bibnamefont{Kovtun}},
  \bibinfo{author}{\bibfnamefont{D.}~\bibnamefont{Son}}, \bibnamefont{and}
  \bibinfo{author}{\bibfnamefont{A.}~\bibnamefont{Starinets}},
  \bibinfo{journal}{Phys.Rev.Lett.} \textbf{\bibinfo{volume}{94}},
  \bibinfo{pages}{111601} (\bibinfo{year}{2005}), \eprint{hep-th/0405231}.

\bibitem[{\citenamefont{Iqbal and Liu}(2009)}]{Iqbal:2008by}
\bibinfo{author}{\bibfnamefont{N.}~\bibnamefont{Iqbal}} \bibnamefont{and}
  \bibinfo{author}{\bibfnamefont{H.}~\bibnamefont{Liu}},
  \bibinfo{journal}{Phys.Rev.} \textbf{\bibinfo{volume}{D79}},
  \bibinfo{pages}{025023} (\bibinfo{year}{2009}), \eprint{0809.3808}.

\bibitem[{\citenamefont{Schafer and Teaney}(2009)}]{Schafer:2009dj}
\bibinfo{author}{\bibfnamefont{T.}~\bibnamefont{Schafer}} \bibnamefont{and}
  \bibinfo{author}{\bibfnamefont{D.}~\bibnamefont{Teaney}},
  \bibinfo{journal}{Rept.Prog.Phys.} \textbf{\bibinfo{volume}{72}},
  \bibinfo{pages}{126001} (\bibinfo{year}{2009}), \eprint{0904.3107}.

\bibitem[{\citenamefont{Heinz and Snellings}(2013)}]{Heinz:2013th}
\bibinfo{author}{\bibfnamefont{U.}~\bibnamefont{Heinz}} \bibnamefont{and}
  \bibinfo{author}{\bibfnamefont{R.}~\bibnamefont{Snellings}},
  \bibinfo{journal}{Ann.Rev.Nucl.Part.Sci.} \textbf{\bibinfo{volume}{63}},
  \bibinfo{pages}{123} (\bibinfo{year}{2013}), \eprint{1301.2826}.

\bibitem[{\citenamefont{Pines and Nozi\`{e}res}(1966)}]{PinesNozieres}
\bibinfo{author}{\bibfnamefont{D.}~\bibnamefont{Pines}} \bibnamefont{and}
  \bibinfo{author}{\bibfnamefont{P.}~\bibnamefont{Nozi\`{e}res}},
  \emph{\bibinfo{title}{{The theory of quantum liquids}}}
  (\bibinfo{publisher}{W.~A.~Benjamin}, \bibinfo{address}{New York},
  \bibinfo{year}{1966}).

\bibitem[{\citenamefont{Lucas et~al.}(2014)\citenamefont{Lucas, Sachdev, and
  Schalm}}]{Lucas:2014zea}
\bibinfo{author}{\bibfnamefont{A.}~\bibnamefont{Lucas}},
  \bibinfo{author}{\bibfnamefont{S.}~\bibnamefont{Sachdev}}, \bibnamefont{and}
  \bibinfo{author}{\bibfnamefont{K.}~\bibnamefont{Schalm}},
  \bibinfo{journal}{Phys.Rev.} \textbf{\bibinfo{volume}{D89}},
  \bibinfo{pages}{066018} (\bibinfo{year}{2014}), \eprint{1401.7993}.

\bibitem[{\citenamefont{Vegh}(2013)}]{Vegh:2013sk}
\bibinfo{author}{\bibfnamefont{D.}~\bibnamefont{Vegh}} (\bibinfo{year}{2013}),
  \eprint{1301.0537}.

\bibitem[{\citenamefont{Forster}(1995)}]{Forster}
\bibinfo{author}{\bibfnamefont{D.}~\bibnamefont{Forster}},
  \emph{\bibinfo{title}{{Hydrodynamic Fluctuations, Broken Symmetry, and
  Correlation Functions}}} (\bibinfo{publisher}{Westview Press},
  \bibinfo{year}{1995}).

\bibitem[{\citenamefont{{Spivak} and {Kivelson}}(2005)}]{2005JPhy4.131..255S}
\bibinfo{author}{\bibfnamefont{B.}~\bibnamefont{{Spivak}}} \bibnamefont{and}
  \bibinfo{author}{\bibfnamefont{S.}~\bibnamefont{{Kivelson}}},
  \bibinfo{journal}{Journal de Physique IV} \textbf{\bibinfo{volume}{131}},
  \bibinfo{pages}{255} (\bibinfo{year}{2005}), \eprint{arXiv:cond-mat/0510422}.

\bibitem[{\citenamefont{Andreev et~al.}(2011)\citenamefont{Andreev, Kivelson,
  and Spivak}}]{PhysRevLett.106.256804}
\bibinfo{author}{\bibfnamefont{A.~V.} \bibnamefont{Andreev}},
  \bibinfo{author}{\bibfnamefont{S.~A.} \bibnamefont{Kivelson}},
  \bibnamefont{and} \bibinfo{author}{\bibfnamefont{B.}~\bibnamefont{Spivak}},
  \bibinfo{journal}{Phys. Rev. Lett.} \textbf{\bibinfo{volume}{106}},
  \bibinfo{pages}{256804} (\bibinfo{year}{2011}).

\bibitem[{\citenamefont{Lawrence and Wilkins}(1973)}]{PhysRevB.7.2317}
\bibinfo{author}{\bibfnamefont{W.~E.} \bibnamefont{Lawrence}} \bibnamefont{and}
  \bibinfo{author}{\bibfnamefont{J.~W.} \bibnamefont{Wilkins}},
  \bibinfo{journal}{Phys. Rev. B} \textbf{\bibinfo{volume}{7}},
  \bibinfo{pages}{2317} (\bibinfo{year}{1973}).

\bibitem[{\citenamefont{Kovtun}(2012)}]{Kovtun:2012rj}
\bibinfo{author}{\bibfnamefont{P.}~\bibnamefont{Kovtun}},
  \bibinfo{journal}{J.Phys.} \textbf{\bibinfo{volume}{A45}},
  \bibinfo{pages}{473001} (\bibinfo{year}{2012}), \eprint{1205.5040}.

\bibitem[{\citenamefont{Balasubramanian and
  Herzog}(2013)}]{Balasubramanian:2013yqa}
\bibinfo{author}{\bibfnamefont{K.}~\bibnamefont{Balasubramanian}}
  \bibnamefont{and} \bibinfo{author}{\bibfnamefont{C.~P.} \bibnamefont{Herzog}}
  (\bibinfo{year}{2013}), \eprint{1312.4953}.

\bibitem[{\citenamefont{Kovtun et~al.}(2011)\citenamefont{Kovtun, Moore, and
  Romatschke}}]{Kovtun:2011np}
\bibinfo{author}{\bibfnamefont{P.}~\bibnamefont{Kovtun}},
  \bibinfo{author}{\bibfnamefont{G.~D.} \bibnamefont{Moore}}, \bibnamefont{and}
  \bibinfo{author}{\bibfnamefont{P.}~\bibnamefont{Romatschke}},
  \bibinfo{journal}{Phys.Rev.} \textbf{\bibinfo{volume}{D84}},
  \bibinfo{pages}{025006} (\bibinfo{year}{2011}), \eprint{1104.1586}.

\bibitem[{\citenamefont{Herzog}(2009)}]{Herzog:2009xv}
\bibinfo{author}{\bibfnamefont{C.~P.} \bibnamefont{Herzog}},
  \bibinfo{journal}{J.Phys.} \textbf{\bibinfo{volume}{A42}},
  \bibinfo{pages}{343001} (\bibinfo{year}{2009}), \eprint{0904.1975}.

\bibitem[{\citenamefont{Alloul et~al.}(2009)\citenamefont{Alloul, Bobroff,
  Gabay, and Hirschfeld}}]{RevModPhys.81.45}
\bibinfo{author}{\bibfnamefont{H.}~\bibnamefont{Alloul}},
  \bibinfo{author}{\bibfnamefont{J.}~\bibnamefont{Bobroff}},
  \bibinfo{author}{\bibfnamefont{M.}~\bibnamefont{Gabay}}, \bibnamefont{and}
  \bibinfo{author}{\bibfnamefont{P.~J.} \bibnamefont{Hirschfeld}},
  \bibinfo{journal}{Rev. Mod. Phys.} \textbf{\bibinfo{volume}{81}},
  \bibinfo{pages}{45} (\bibinfo{year}{2009}).

\bibitem[{\citenamefont{McGreevy}(2010)}]{McGreevy:2009xe}
\bibinfo{author}{\bibfnamefont{J.}~\bibnamefont{McGreevy}},
  \bibinfo{journal}{Adv.High Energy Phys.} \textbf{\bibinfo{volume}{2010}},
  \bibinfo{pages}{723105} (\bibinfo{year}{2010}), \eprint{0909.0518}.

\bibitem[{\citenamefont{Hartnoll}(2009)}]{Hartnoll:2009sz}
\bibinfo{author}{\bibfnamefont{S.~A.} \bibnamefont{Hartnoll}},
  \bibinfo{journal}{Class.Quant.Grav.} \textbf{\bibinfo{volume}{26}},
  \bibinfo{pages}{224002} (\bibinfo{year}{2009}), \eprint{0903.3246}.

\bibitem[{\citenamefont{Hartnoll}(2011)}]{Hartnoll:2011fn}
\bibinfo{author}{\bibfnamefont{S.~A.} \bibnamefont{Hartnoll}}
  (\bibinfo{year}{2011}), \eprint{1106.4324}.

\bibitem[{\citenamefont{Gubser}(2008)}]{Gubser:2008px}
\bibinfo{author}{\bibfnamefont{S.~S.} \bibnamefont{Gubser}},
  \bibinfo{journal}{Phys.Rev.} \textbf{\bibinfo{volume}{D78}},
  \bibinfo{pages}{065034} (\bibinfo{year}{2008}), \eprint{0801.2977}.

\bibitem[{\citenamefont{Hartnoll et~al.}(2008)\citenamefont{Hartnoll, Herzog,
  and Horowitz}}]{Hartnoll:2008vx}
\bibinfo{author}{\bibfnamefont{S.~A.} \bibnamefont{Hartnoll}},
  \bibinfo{author}{\bibfnamefont{C.~P.} \bibnamefont{Herzog}},
  \bibnamefont{and} \bibinfo{author}{\bibfnamefont{G.~T.}
  \bibnamefont{Horowitz}}, \bibinfo{journal}{Phys.Rev.Lett.}
  \textbf{\bibinfo{volume}{101}}, \bibinfo{pages}{031601}
  (\bibinfo{year}{2008}), \eprint{0803.3295}.

\bibitem[{\citenamefont{Liu et~al.}(2011)\citenamefont{Liu, McGreevy, and
  Vegh}}]{Liu:2009dm}
\bibinfo{author}{\bibfnamefont{H.}~\bibnamefont{Liu}},
  \bibinfo{author}{\bibfnamefont{J.}~\bibnamefont{McGreevy}}, \bibnamefont{and}
  \bibinfo{author}{\bibfnamefont{D.}~\bibnamefont{Vegh}},
  \bibinfo{journal}{Phys.Rev.} \textbf{\bibinfo{volume}{D83}},
  \bibinfo{pages}{065029} (\bibinfo{year}{2011}), \eprint{0903.2477}.

\bibitem[{\citenamefont{Cubrovic et~al.}(2009)\citenamefont{Cubrovic, Zaanen,
  and Schalm}}]{Cubrovic:2009ye}
\bibinfo{author}{\bibfnamefont{M.}~\bibnamefont{Cubrovic}},
  \bibinfo{author}{\bibfnamefont{J.}~\bibnamefont{Zaanen}}, \bibnamefont{and}
  \bibinfo{author}{\bibfnamefont{K.}~\bibnamefont{Schalm}},
  \bibinfo{journal}{Science} \textbf{\bibinfo{volume}{325}},
  \bibinfo{pages}{439} (\bibinfo{year}{2009}), \eprint{0904.1993}.

\bibitem[{\citenamefont{Faulkner et~al.}(2011)\citenamefont{Faulkner, Liu,
  McGreevy, and Vegh}}]{Faulkner:2009wj}
\bibinfo{author}{\bibfnamefont{T.}~\bibnamefont{Faulkner}},
  \bibinfo{author}{\bibfnamefont{H.}~\bibnamefont{Liu}},
  \bibinfo{author}{\bibfnamefont{J.}~\bibnamefont{McGreevy}}, \bibnamefont{and}
  \bibinfo{author}{\bibfnamefont{D.}~\bibnamefont{Vegh}},
  \bibinfo{journal}{Phys.Rev.} \textbf{\bibinfo{volume}{D83}},
  \bibinfo{pages}{125002} (\bibinfo{year}{2011}), \eprint{0907.2694}.

\bibitem[{\citenamefont{Gubser and Rocha}(2010)}]{Gubser:2009qt}
\bibinfo{author}{\bibfnamefont{S.~S.} \bibnamefont{Gubser}} \bibnamefont{and}
  \bibinfo{author}{\bibfnamefont{F.~D.} \bibnamefont{Rocha}},
  \bibinfo{journal}{Phys.Rev.} \textbf{\bibinfo{volume}{D81}},
  \bibinfo{pages}{046001} (\bibinfo{year}{2010}), \eprint{0911.2898}.

\bibitem[{\citenamefont{Horowitz
  et~al.}(2012{\natexlab{a}})\citenamefont{Horowitz, Santos, and
  Tong}}]{Horowitz:2012ky}
\bibinfo{author}{\bibfnamefont{G.~T.} \bibnamefont{Horowitz}},
  \bibinfo{author}{\bibfnamefont{J.~E.} \bibnamefont{Santos}},
  \bibnamefont{and} \bibinfo{author}{\bibfnamefont{D.}~\bibnamefont{Tong}},
  \bibinfo{journal}{JHEP} \textbf{\bibinfo{volume}{1207}}, \bibinfo{pages}{168}
  (\bibinfo{year}{2012}{\natexlab{a}}), \eprint{1204.0519}.

\bibitem[{\citenamefont{Horowitz
  et~al.}(2012{\natexlab{b}})\citenamefont{Horowitz, Santos, and
  Tong}}]{Horowitz:2012gs}
\bibinfo{author}{\bibfnamefont{G.~T.} \bibnamefont{Horowitz}},
  \bibinfo{author}{\bibfnamefont{J.~E.} \bibnamefont{Santos}},
  \bibnamefont{and} \bibinfo{author}{\bibfnamefont{D.}~\bibnamefont{Tong}},
  \bibinfo{journal}{JHEP} \textbf{\bibinfo{volume}{1211}}, \bibinfo{pages}{102}
  (\bibinfo{year}{2012}{\natexlab{b}}), \eprint{1209.1098}.

\bibitem[{\citenamefont{Ling et~al.}(2013)\citenamefont{Ling, Niu, Wu, and
  Xian}}]{Ling:2013nxa}
\bibinfo{author}{\bibfnamefont{Y.}~\bibnamefont{Ling}},
  \bibinfo{author}{\bibfnamefont{C.}~\bibnamefont{Niu}},
  \bibinfo{author}{\bibfnamefont{J.-P.} \bibnamefont{Wu}}, \bibnamefont{and}
  \bibinfo{author}{\bibfnamefont{Z.-Y.} \bibnamefont{Xian}}
  (\bibinfo{year}{2013}), \eprint{1309.4580}.

\bibitem[{\citenamefont{Cvetic et~al.}(1999)\citenamefont{Cvetic, Duff, Hoxha,
  Liu, Lu et~al.}}]{Cvetic:1999xp}
\bibinfo{author}{\bibfnamefont{M.}~\bibnamefont{Cvetic}},
  \bibinfo{author}{\bibfnamefont{M.}~\bibnamefont{Duff}},
  \bibinfo{author}{\bibfnamefont{P.}~\bibnamefont{Hoxha}},
  \bibinfo{author}{\bibfnamefont{J.~T.} \bibnamefont{Liu}},
  \bibinfo{author}{\bibfnamefont{H.}~\bibnamefont{Lu}}, \bibnamefont{et~al.},
  \bibinfo{journal}{Nucl.Phys.} \textbf{\bibinfo{volume}{B558}},
  \bibinfo{pages}{96} (\bibinfo{year}{1999}), \eprint{hep-th/9903214}.

\bibitem[{\citenamefont{Iqbal et~al.}(2012)\citenamefont{Iqbal, Liu, and
  Mezei}}]{Iqbal:2011in}
\bibinfo{author}{\bibfnamefont{N.}~\bibnamefont{Iqbal}},
  \bibinfo{author}{\bibfnamefont{H.}~\bibnamefont{Liu}}, \bibnamefont{and}
  \bibinfo{author}{\bibfnamefont{M.}~\bibnamefont{Mezei}},
  \bibinfo{journal}{JHEP} \textbf{\bibinfo{volume}{1204}}, \bibinfo{pages}{086}
  (\bibinfo{year}{2012}), \eprint{1105.4621}.

\bibitem[{\citenamefont{Goldstein et~al.}(2010)\citenamefont{Goldstein, Kachru,
  Prakash, and Trivedi}}]{Goldstein:2009cv}
\bibinfo{author}{\bibfnamefont{K.}~\bibnamefont{Goldstein}},
  \bibinfo{author}{\bibfnamefont{S.}~\bibnamefont{Kachru}},
  \bibinfo{author}{\bibfnamefont{S.}~\bibnamefont{Prakash}}, \bibnamefont{and}
  \bibinfo{author}{\bibfnamefont{S.~P.} \bibnamefont{Trivedi}},
  \bibinfo{journal}{JHEP} \textbf{\bibinfo{volume}{1008}}, \bibinfo{pages}{078}
  (\bibinfo{year}{2010}), \eprint{0911.3586}.

\bibitem[{\citenamefont{Charmousis et~al.}(2010)\citenamefont{Charmousis,
  Gouteraux, Kim, Kiritsis, and Meyer}}]{Charmousis:2010zz}
\bibinfo{author}{\bibfnamefont{C.}~\bibnamefont{Charmousis}},
  \bibinfo{author}{\bibfnamefont{B.}~\bibnamefont{Gouteraux}},
  \bibinfo{author}{\bibfnamefont{B.}~\bibnamefont{Kim}},
  \bibinfo{author}{\bibfnamefont{E.}~\bibnamefont{Kiritsis}}, \bibnamefont{and}
  \bibinfo{author}{\bibfnamefont{R.}~\bibnamefont{Meyer}},
  \bibinfo{journal}{JHEP} \textbf{\bibinfo{volume}{1011}}, \bibinfo{pages}{151}
  (\bibinfo{year}{2010}), \eprint{1005.4690}.

\bibitem[{\citenamefont{Hartnoll and Shaghoulian}(2012)}]{Hartnoll:2012wm}
\bibinfo{author}{\bibfnamefont{S.~A.} \bibnamefont{Hartnoll}} \bibnamefont{and}
  \bibinfo{author}{\bibfnamefont{E.}~\bibnamefont{Shaghoulian}},
  \bibinfo{journal}{JHEP} \textbf{\bibinfo{volume}{1207}}, \bibinfo{pages}{078}
  (\bibinfo{year}{2012}), \eprint{1203.4236}.

\bibitem[{\citenamefont{Davison and Parnachev}(2013)}]{Davison:2013bxa}
\bibinfo{author}{\bibfnamefont{R.~A.} \bibnamefont{Davison}} \bibnamefont{and}
  \bibinfo{author}{\bibfnamefont{A.}~\bibnamefont{Parnachev}},
  \bibinfo{journal}{JHEP} \textbf{\bibinfo{volume}{1306}}, \bibinfo{pages}{100}
  (\bibinfo{year}{2013}), \eprint{1303.6334}.

\bibitem[{\citenamefont{Davison et~al.}(2013)\citenamefont{Davison, Goykhman,
  and Parnachev}}]{Davison:2013uha}
\bibinfo{author}{\bibfnamefont{R.~A.} \bibnamefont{Davison}},
  \bibinfo{author}{\bibfnamefont{M.}~\bibnamefont{Goykhman}}, \bibnamefont{and}
  \bibinfo{author}{\bibfnamefont{A.}~\bibnamefont{Parnachev}}
  (\bibinfo{year}{2013}), \eprint{1312.0463}.

\bibitem[{\citenamefont{Tarrio}(2014)}]{Tarrio:2013tta}
\bibinfo{author}{\bibfnamefont{J.}~\bibnamefont{Tarrio}},
  \bibinfo{journal}{JHEP} \textbf{\bibinfo{volume}{1404}}, \bibinfo{pages}{042}
  (\bibinfo{year}{2014}), \eprint{1312.2902}.

\bibitem[{\citenamefont{Davison}(2013)}]{Davison:2013jba}
\bibinfo{author}{\bibfnamefont{R.~A.} \bibnamefont{Davison}},
  \bibinfo{journal}{Phys.Rev.} \textbf{\bibinfo{volume}{D88}},
  \bibinfo{pages}{086003} (\bibinfo{year}{2013}), \eprint{1306.5792}.

\bibitem[{\citenamefont{Blake and Tong}(2013)}]{Blake:2013bqa}
\bibinfo{author}{\bibfnamefont{M.}~\bibnamefont{Blake}} \bibnamefont{and}
  \bibinfo{author}{\bibfnamefont{D.}~\bibnamefont{Tong}},
  \bibinfo{journal}{Phys.Rev.} \textbf{\bibinfo{volume}{D88}},
  \bibinfo{pages}{106004} (\bibinfo{year}{2013}), \eprint{1308.4970}.

\bibitem[{\citenamefont{Andrade and Withers}(2013)}]{Andrade:2013gsa}
\bibinfo{author}{\bibfnamefont{T.}~\bibnamefont{Andrade}} \bibnamefont{and}
  \bibinfo{author}{\bibfnamefont{B.}~\bibnamefont{Withers}}
  (\bibinfo{year}{2013}), \eprint{1311.5157}.

\bibitem[{\citenamefont{Blake et~al.}(2014)\citenamefont{Blake, Tong, and
  Vegh}}]{Blake:2013owa}
\bibinfo{author}{\bibfnamefont{M.}~\bibnamefont{Blake}},
  \bibinfo{author}{\bibfnamefont{D.}~\bibnamefont{Tong}}, \bibnamefont{and}
  \bibinfo{author}{\bibfnamefont{D.}~\bibnamefont{Vegh}},
  \bibinfo{journal}{Phys.Rev.Lett.} \textbf{\bibinfo{volume}{112}},
  \bibinfo{pages}{071602} (\bibinfo{year}{2014}), \eprint{1310.3832}.

\bibitem[{\citenamefont{Gouteraux}(2014)}]{Gouteraux:2014hca}
\bibinfo{author}{\bibfnamefont{B.}~\bibnamefont{Gouteraux}}
  (\bibinfo{year}{2014}), \eprint{1401.5436}.

\bibitem[{\citenamefont{Donos and Gauntlett}(2014)}]{Donos:2014uba}
\bibinfo{author}{\bibfnamefont{A.}~\bibnamefont{Donos}} \bibnamefont{and}
  \bibinfo{author}{\bibfnamefont{J.~P.} \bibnamefont{Gauntlett}}
  (\bibinfo{year}{2014}), \eprint{1401.5077}.

\bibitem[{\citenamefont{Loram et~al.}(1993)\citenamefont{Loram, Mirza, Cooper,
  and Liang}}]{PhysRevLett.71.1740}
\bibinfo{author}{\bibfnamefont{J.~W.} \bibnamefont{Loram}},
  \bibinfo{author}{\bibfnamefont{K.~A.} \bibnamefont{Mirza}},
  \bibinfo{author}{\bibfnamefont{J.~R.} \bibnamefont{Cooper}},
  \bibnamefont{and} \bibinfo{author}{\bibfnamefont{W.~Y.} \bibnamefont{Liang}},
  \bibinfo{journal}{Phys. Rev. Lett.} \textbf{\bibinfo{volume}{71}},
  \bibinfo{pages}{1740} (\bibinfo{year}{1993}).

\bibitem[{\citenamefont{{Barisic} et~al.}(2013)\citenamefont{{Barisic}, {Chan},
  {Li}, {Yu}, {Zhao} et~al.}}]{2013PNAS..11012235B}
\bibinfo{author}{\bibfnamefont{N.}~\bibnamefont{{Barisic}}},
  \bibinfo{author}{\bibfnamefont{M.~K.} \bibnamefont{{Chan}}},
  \bibinfo{author}{\bibfnamefont{Y.}~\bibnamefont{{Li}}},
  \bibinfo{author}{\bibfnamefont{G.}~\bibnamefont{{Yu}}},
  \bibinfo{author}{\bibfnamefont{X.}~\bibnamefont{{Zhao}}},
  \bibnamefont{et~al.}, \bibinfo{journal}{PNAS} \textbf{\bibinfo{volume}{110}},
  \bibinfo{pages}{12235} (\bibinfo{year}{2013}), \eprint{1207.1504}.

\bibitem[{\citenamefont{van~der Marel et~al.}(2003)\citenamefont{van~der Marel,
  Molegraaf, Zaanen, Nussinov, Carbone et~al.}}]{vdmarelpowerlawcuprates}
\bibinfo{author}{\bibfnamefont{D.}~\bibnamefont{van~der Marel}},
  \bibinfo{author}{\bibfnamefont{H.}~\bibnamefont{Molegraaf}},
  \bibinfo{author}{\bibfnamefont{J.}~\bibnamefont{Zaanen}},
  \bibinfo{author}{\bibfnamefont{Z.}~\bibnamefont{Nussinov}},
  \bibinfo{author}{\bibfnamefont{F.}~\bibnamefont{Carbone}},
  \bibnamefont{et~al.}, \bibinfo{journal}{Nature}
  \textbf{\bibinfo{volume}{425}}, \bibinfo{pages}{271} (\bibinfo{year}{2003}),
  \eprint{cond-mat/0309172}.

\bibitem[{\citenamefont{{Mirzaei} et~al.}(2013)\citenamefont{{Mirzaei},
  {Stricker}, {Hancock}, {Berthod}, {Georges} et~al.}}]{2013PNAS..110.5774M}
\bibinfo{author}{\bibfnamefont{S.~I.} \bibnamefont{{Mirzaei}}},
  \bibinfo{author}{\bibfnamefont{D.}~\bibnamefont{{Stricker}}},
  \bibinfo{author}{\bibfnamefont{J.~N.} \bibnamefont{{Hancock}}},
  \bibinfo{author}{\bibfnamefont{C.}~\bibnamefont{{Berthod}}},
  \bibinfo{author}{\bibfnamefont{A.}~\bibnamefont{{Georges}}},
  \bibnamefont{et~al.}, \bibinfo{journal}{PNAS} \textbf{\bibinfo{volume}{110}},
  \bibinfo{pages}{5774} (\bibinfo{year}{2013}), \eprint{1207.6704}.

\bibitem[{\citenamefont{{Shekhter} et~al.}(2013)\citenamefont{{Shekhter},
  {Ramshaw}, {Liang}, {Hardy}, {Bonn} et~al.}}]{2013Natur.498...75S}
\bibinfo{author}{\bibfnamefont{A.}~\bibnamefont{{Shekhter}}},
  \bibinfo{author}{\bibfnamefont{B.~J.} \bibnamefont{{Ramshaw}}},
  \bibinfo{author}{\bibfnamefont{R.}~\bibnamefont{{Liang}}},
  \bibinfo{author}{\bibfnamefont{W.~N.} \bibnamefont{{Hardy}}},
  \bibinfo{author}{\bibfnamefont{D.~A.} \bibnamefont{{Bonn}}},
  \bibnamefont{et~al.}, \bibinfo{journal}{\nat} \textbf{\bibinfo{volume}{498}},
  \bibinfo{pages}{75} (\bibinfo{year}{2013}), \eprint{1208.5810}.

\bibitem[{\citenamefont{Chien et~al.}(1991)\citenamefont{Chien, Wang, and
  Ong}}]{PhysRevLett.67.2088}
\bibinfo{author}{\bibfnamefont{T.~R.} \bibnamefont{Chien}},
  \bibinfo{author}{\bibfnamefont{Z.~Z.} \bibnamefont{Wang}}, \bibnamefont{and}
  \bibinfo{author}{\bibfnamefont{N.~P.} \bibnamefont{Ong}},
  \bibinfo{journal}{Phys. Rev. Lett.} \textbf{\bibinfo{volume}{67}},
  \bibinfo{pages}{2088} (\bibinfo{year}{1991}).

\bibitem[{\citenamefont{Anderson}(1991)}]{PhysRevLett.67.2092}
\bibinfo{author}{\bibfnamefont{P.~W.} \bibnamefont{Anderson}},
  \bibinfo{journal}{Phys. Rev. Lett.} \textbf{\bibinfo{volume}{67}},
  \bibinfo{pages}{2092} (\bibinfo{year}{1991}).

\bibitem[{\citenamefont{{Bruin} et~al.}(2013)\citenamefont{{Bruin}, {Sakai},
  {Perry}, and {Mackenzie}}}]{2013Sci...339..804B}
\bibinfo{author}{\bibfnamefont{J.~A.~N.} \bibnamefont{{Bruin}}},
  \bibinfo{author}{\bibfnamefont{H.}~\bibnamefont{{Sakai}}},
  \bibinfo{author}{\bibfnamefont{R.~S.} \bibnamefont{{Perry}}},
  \bibnamefont{and} \bibinfo{author}{\bibfnamefont{A.~P.}
  \bibnamefont{{Mackenzie}}}, \bibinfo{journal}{Science}
  \textbf{\bibinfo{volume}{339}}, \bibinfo{pages}{804} (\bibinfo{year}{2013}).

\end{thebibliography}

\end{document}